\documentclass[aps,prd,nofootinbib,showpacs,eqsecnum,superscriptaddress,longbibliography,preprintnumbers]{revtex4-2}
\usepackage{amsmath, dsfont}
\allowdisplaybreaks 
\usepackage{amssymb}
\usepackage{epsfig}
\usepackage{calc}
\usepackage{amsfonts}
\usepackage{mathtools}
\usepackage{enumerate}
\usepackage{graphicx}
\usepackage[ansinew]{inputenc}
\usepackage{multirow}
\usepackage{subcaption}
\usepackage{tensor}

\usepackage[dvipsnames]{xcolor}
\usepackage{natbib}
\usepackage{cancel}
\usepackage{soul}

\usepackage{etoolbox}
\makeatletter
\patchcmd{\@bibsetup}{\raggedright}{}{}{}
\makeatother

\usepackage{setspace}
\usepackage{empheq} 
\usepackage[skins]{tcolorbox}

\graphicspath{{Figs/}}

\usepackage{libertine}

\definecolor{azure(colorwheel)}{rgb}{0.0, 0.5, 1.0}
\definecolor{PRDblue}{RGB}{48,46,146}
\definecolor{nicered}{rgb}{0.7,0.1,0.1}
\definecolor{DarkViolet}{RGB}{148,0,211}
\definecolor{myDarkBlue}{rgb}{0,0.1,0.7}
\definecolor{tealgreen}{rgb}{0.0, 0.51, 0.5}

\usepackage{hyperref}
\hypersetup{colorlinks,
    bookmarksopen,
	bookmarksnumbered,
	citecolor={nicered},
	linkcolor={PRDblue},
	urlcolor={tealgreen},
	pdfstartview=FitH}

\newcommand*\diff{\mathop{}\!\mathrm{d}}

\usepackage{orcidlink}
\def\idbako{\orcidlink{0000-0002-3012-6144}}
\def\idchar{\orcidlink{0000-0002-5364-4753}}
\def\idchat{\orcidlink{0000-0003-4479-2970}}
\def\idnakas{\orcidlink{0000-0002-3522-5803}}

\begin{document}

\tcbset{frame style={top color=gray!20,
                     bottom color=gray!20},
        fonttitle=\bfseries,coltitle=black, sharp corners,boxrule=0pt}

\setcounter{page}{1}
\title[]{\fontsize{18}{18}\selectfont{Theoretical filters for shift-symmetric Horndeski gravities}}

\author{Athanasios Bakopoulos\,\idbako}
\email{atbakopoulos@gmail.com}
\affiliation{Division of Applied Analysis, Department of Mathematics, University of Patras, Rio Patras GR-26504, Greece}
\affiliation{Physics Division, School of Applied Mathematical and Physical Sciences, National Technical University of Athens, Zografou Campus, Athens 15780, Greece}
\author{Christos Charmousis\,\idchar}
\email{christos@ijclab.in2p3.fr}
\affiliation{Universit\'e Paris-Saclay, CNRS/IN2P3, IJCLab, 91405 Orsay, France}
\author{Nikos Chatzifotis\,\idchat}
\email{chatzifotisn@gmail.com}
\affiliation{Physics Division, School of Applied Mathematical and Physical Sciences, National Technical University of Athens, Zografou Campus, Athens 15780, Greece}
\author{Theodoros Nakas\,\idnakas}
\email{theodoros.nakas@gmail.com}
\affiliation{Cosmology, Gravity, and Astroparticle Physics Group, Center for Theoretical Physics of the Universe, Institute for Basic Science (IBS), Daejeon 34126, Korea}

\begin{abstract}
    \noindent\textbf{Abstract.} We investigate the structure of nontrivial maximally symmetric vacua and compact-object solutions in shift-symmetric scalar-tensor theories. Focusing on Horndeski gravity, we derive consistency conditions directly from the field equations to identify the subclasses that admit Minkowski and de Sitter vacua with a nontrivial scalar field. In doing so, we obtain a filtering mechanism that operates independently of observational data. In this context, we introduce the notion of stealth vacua, where the scalar field remains active without altering the vacuum. Following this, we examine the theoretical framework of Horndeski theories that admit homogeneous geometries and we extract the implicit form of the solution pertaining to the entire family of theories. Building upon these frameworks, we construct exact solutions in beyond-Horndeski gravity by applying a linear disformal transformation to the regularized Einstein-Gauss-Bonnet black hole. This procedure yields solitonic spacetimes with scalar hair as well as black holes carrying primary scalar hair, demonstrating how disformal maps can qualitatively modify solution properties. We delineate the parameter space in which the transformation is well-defined and analyze the solutions. Our results provide both a principled criterion for selecting viable Horndeski models and a framework for exploring rich solution spaces in beyond-Horndeski gravity.
\end{abstract}
\maketitle
\tableofcontents

\section{Introduction}

Over the past century, General Relativity (GR) has established itself as the standard theory of gravitation, successfully accounting for a wide range of phenomena from the precession of Mercury's perihelion to the recent detection of gravitational waves (GW) \cite{LIGOScientific:2016aoc}. 
Despite this remarkable success, GR is widely regarded as incomplete. Persistent open questions, such as the nature of dark energy \cite{Nojiri:2017ncd, Bahamonde:2017ize, DES:2021wwk, DESI:2024mwx, Langlois:2018dxi, vandeBruck:2025aaa}, the cosmological constant problem and other cosmological tensions \cite{Perivolaropoulos:2021jda, Abdalla:2022yfr, CosmoVerseNetwork:2025alb}, the resolution of curvature singularities, and the ultraviolet completion of the theory \cite{Stelle:1976gc, Niedermaier:2006wt}, suggest that GR may represent only an effective description of gravity at accessible scales. 
These challenges have motivated the exploration of alternative and extended theories of gravity \cite{Horndeski:1974wa, Boulware:1985wk, Randall:1999ee, Randall:1999vf, Nojiri:2010wj, deRham:2010kj, Kobayashi:2010cm, Charmousis:2011bf, Charmousis:2011ea, Capozziello:2011et,  Gleyzes:2014dya, Langlois:2015cwa, Cai:2015emx, Kobayashi:2019hrl, Bahamonde:2021gfp, Charmousis:2025jpx}, aiming either to provide a more fundamental framework or to capture deviations that could become observable in high-energy or strong-field regimes.

Among the various candidates, scalar-tensor (ST) theories have emerged as particularly promising extensions. 
By augmenting the gravitational sector with an additional scalar degree of freedom, they open up a wide and flexible landscape of possibilities while maintaining a close connection to GR. 
Within this class, Horndeski gravity \cite{Horndeski:1974wa} plays a central role, as it is the most general scalar-tensor theory in four dimensions that yields second-order equations of motion for both the metric and the scalar field. 
The price of this generality is the large functional freedom of the theory, which makes the classification of physically meaningful models a nontrivial task. 
A fruitful simplification is obtained by imposing shift symmetry on the scalar, $\phi\rightarrow\phi+\rm{const.}$, which constrains the allowed interactions to depend only on derivatives of the scalar field. This symmetry not only reduces the complexity of the theory but also leads to technically and physically interesting features, including the conservation of a Noether current and the possibility of nontrivial scalar profiles compatible with static and spherically symmetric geometries. Such theories admit a $\phi=$constant trivial vacuum for which we have the standard maximally symmetric GR vacua of constant curvature, while they also support a linear temporal dependence of the scalar field without spoiling the stationarity of the underlying geometry.
In particular, scalar fields of the form $\phi=qt+\psi(r)$ provide a robust mechanism for generating solutions where the scalar remains nontrivial even in vacuum spacetimes.

These scalar configurations have played a central role in the discovery of novel solutions in shift-symmetric Horndeski theories for static \cite{Babichev:2012re, Babichev:2013cya, Kobayashi:2014eva, Motohashi:2019sen,
Bakopoulos:2023fmv, Baake:2023zsq, Bakopoulos:2023sdm} as well as for rotating backgrounds \cite{Charmousis:2019vnf, Anson:2020trg, BenAchour:2020fgy, Babichev:2023mgk}. 
Early studies, initiated in \cite{Babichev:2013cya}, revealed the existence of stealth black holes \cite{Charmousis:2015aya, Minamitsuji:2018vuw, BenAchour:2018dap, Takahashi:2020hso}, in which the geometry coincides with a standard GR solution, such as Schwarzschild or Schwarzschild-de Sitter, while the scalar field develops a nontrivial profile that leaves the metric unaffected. 
More recently however, it has been shown that certain subclasses of shift-symmetric Horndeski and beyond Horndeski gravity admit black-hole solutions endowed with primary scalar hair \cite{Bakopoulos:2023fmv, Baake:2023zsq, Bakopoulos:2023sdm,Charmousis:2025xug}. 
The primary hair in question is present due to the specific linear time dependence of the scalar field, ie. due to the nontrivial time dependence of the scalar which in turn stems from the global shift symmetry.
In this case, the scalar field carries an independent integration constant that is not fixed by the mass or other geometric parameters of the spacetime, in sharp contrast with the traditional notion of secondary hair. 
The presence of primary hair represents a fundamental extension of the spectrum of gravitational solutions, as it introduces new independent charges associated with the scalar sector and enriches the phenomenology of compact objects. 
These developments highlight that, within the restricted class of shift-symmetric models, the interplay between the scalar degree of freedom and the geometry gives a much broader set of physically relevant solutions than those available in GR.

In General Relativity, the concept of vacuum solutions is unique and plays a central role in the structure of the theory. 
Within the framework of GR, the term ``vacuum solution" will denote any spacetime that satisfies the Einstein equations in the complete absence of matter throughout the entire manifold.
Consequently, Minkowski and de Sitter geometries stand out as the maximally symmetric and physically significant vacua in four dimensions. 
Minkowski spacetime represents the flat, static background underlying local physics and serves as the limiting case of more complex spacetimes when all characteristic scales vanish (such as mass, electric or magnetic charge).
Similarly, de Sitter spacetime provides the maximally symmetric configuration with positive constant curvature, capturing the physics of a universe dominated by a cosmological constant. 
These two geometries are not only exact solutions but also constitute the reference points around which perturbative analysis is constructed for spacetimes with lesser symmetry.
For instance, gravitational waves are defined as perturbations propagating on a Minkowski or de Sitter background, while certain cosmological fluctuations are studied by expanding around the de Sitter solution. 
Beyond their role in perturbation theory, they also function as the asymptotic limits of more complicated configurations, such as Schwarzschild(-de Sitter) and Kerr(-de Sitter) spacetimes.
Therefore, Minkowski and de Sitter vacua provide the backbone of GR solutions, acting as the simplest yet most universal geometries upon which the full richness of gravitational physics is built.

In scalar-tensor theories such as Horndeski gravity, the notion of maximally symmetric vacua turns out to be nontrivial especially if we are seeking hairy solutions connected to the vacua in question. In this paper, we will seek all the shift symmetric Horndeski theories that admit nontrivial (in the scalar field) maximally symmetric vacua with positive or zero cosmological constant. It turns out that this theoretical filter already constraints quite severely the theories in question. The idea is that within these theories hairy solutions may exist, and when they do, they will be asymptotically connected to the aforementioned maximally symmetric vacua. In this sense, the nontrivial vacuum continues to serve as the true background spacetime, the stage on which gravitational waves propagate or cosmological perturbations are defined, just as Minkowski and de Sitter space do in General Relativity. With our analysis we introduce for the first time the notion of stealth vacua configurations where the scalar field remains active while leaving no imprint on the vacuum metric.

Our approach to this problem is to impose theoretical consistency conditions directly at the level of the field equations, rather than relying on observational constraints or phenomenological arguments. 
We ask a basic yet stringent question: which shift-symmetric Horndeski theories are capable of supporting nontrivial vacuum solutions, ie., maximally symmetric positive or zero curvature spacetimes with a nontrivial scalar field. It is within these theories that nontrivial, ie., non GR black-hole solutions may exist.
This perspective allows us to constrain the theory space from first principles, by demanding the existence of backgrounds that any physically reasonable model of gravity must accommodate. 
In doing so, we obtain a filtering mechanism that operates independently of observational data. 
The viability of a theory is determined not by comparison with a particular astrophysical environment, but by its ability to sustain the fundamental vacua that underpin the dynamics of gravitational perturbations. 
This provides a clean and robust criterion for model selection. 
Theories that pass this test retain the essential structural elements required of a consistent gravitational framework, while those that fail are excluded from further consideration. 
The strength of this strategy lies in its generality and clarity: it provides unambiguous selection rules that remain valid regardless of empirical uncertainties, and it identifies a restricted subset of Horndeski models that are theoretically admissible at the most basic level. 
In this sense, our analysis identifies the minimal conditions that any extended theory of gravity must fulfill to play the same foundational role as GR. 
Beyond the immediate results, this approach also establishes a framework that can be applied more broadly.
By treating the existence of vacua as an a priori starting point, one gains a principled and systematic tool for navigating the vast landscape of scalar-tensor modifications of gravity.

Having established the criteria that single out Horndeski subclasses, we turn to the role of disformal transformations as a means of constructing exact solutions in beyond-Horndeski gravity. In particular, using the regularized Einstein-Gauss-Bonnet theory \cite{Lu:2020iav, Fernandes:2020nbq, Hennigar:2020lsl, Fernandes:2021dsb} and the black hole solution found in \cite{Charmousis:2019vnf} as our seed solution in Horndeski gravity, we construct exact nonhomogeneous compact-object solutions, including black holes, in the corresponding beyond-Horndeski theory.
An important feature of this procedure is that seed configurations with well-defined vacua can be mapped to new geometries that inherit these vacuum properties, ensuring consistency across different scalar-tensor frameworks.
We first demonstrate that applying this transformation to the Minkowski background with a nontrivial scalar field in Horndeski gravity yields a solitonic solution endowed with scalar hair in the target theory. 
Then, we extend this procedure to the full black-hole solution and find that the disformal map generates black holes with primary scalar hair in beyond-Horndeski gravity \cite{Bakopoulos:2023fmv, Baake:2023zsq, Bakopoulos:2023sdm, Bakopoulos:2024ogt, Myung:2025afs}. 
Finally, we analyze the parameter space in which the transformation is valid and examine the properties of the resulting solutions.

The structure of the paper is as follows: In Section \eqref{thfram}, we derive and express in a convenient form the field equations of the general shift-symmetric Horndeski gravity assuming a spherically symmetric and static spacetime.
Then, we find the theory constraints for nontrivial Minkowski vacua accompanied by the scalar field solution in question in Section \eqref{mink}. Within these Horndeski theories we then treat two example cases where we have a non zero ADM mass \cite{Arnowitt:1959ah} resulting in a homogenous and a nonhomogeneous spacetime solution. We end the section discussing several special cases. In Section \eqref{yo}, we establish the theories admitting homogeneous black-hole solutions and give an implicit solution for all such theories. We then turn to de Sitter vacua in Section \eqref{yo1} and end with the cases of all $X$ constant solutions in Section \eqref{yo2}. In the last Section \eqref{yo3}, we discuss the disformal transformations of a particular explicit black-hole solution constructing in this way black-hole spacetimes with primary hair. In the final section we give some concluding remarks.


\section{Theoretical framework}
\label{thfram}

We consider the shift symmetric ($\phi\to \phi + {\rm const.}$) Horndeski theory which is described by the action 
\begin{equation}
    S=\frac{1}{16\pi}\int \diff^4x \sqrt{-g} \left\{\mathcal{L}_2+\mathcal{L}_3+\mathcal{L}_4+\mathcal{L}_5\right\}\,,
\end{equation}
where the Lagrangian densities are given by
\begin{align}
    &\mathcal{L}_2 = G_2(X)\,,\\[2mm]
    &\mathcal{L}_3 = -G_3(X)\square \phi\,,\\[2mm]
    &\mathcal{L}_4 = G_4(X)R + G_{4X}\left[ (\square\phi)^2-(\nabla_\mu \nabla_\nu\phi)^2 \right]\,,\\[2mm]
    &\mathcal{L}_5 = G_5(X)G_{\mu\nu}\nabla^\mu\nabla^\nu\phi-\frac{G_{5X}}{6}\left[ (\square\phi)^3-3(\square\phi) (\nabla_\mu\nabla_\nu\phi)^2+2(\nabla_\mu\nabla_\nu\phi)^3 \right]\,.
\end{align}
Notice that there is no explicit $\phi$-dependence in the model functions of the theory.
Here, $X = -\frac{1}{2} g^{\mu\nu} \partial_{\mu} \phi \partial_{\nu} \phi$ represents the kinetic term of the scalar field $\phi$, while $(\nabla_\mu\nabla_\nu\phi)^2\equiv (\nabla_\mu\nabla_\nu\phi) (\nabla^\mu\nabla^\nu\phi)$ and $(\nabla_\mu\nabla_\nu\phi)^3\equiv (\nabla_\mu\nabla_\nu\phi) (\nabla^\nu\nabla^\lambda\phi)(\nabla_\lambda \nabla^\mu \phi)$.
Note also that in geometrized units ($c=G=1$), the model functions possess the following dimensions: $[G_2]=({\rm length})^{-2}$, $[G_3]=[G_4]=({\rm dimensionless})$, and $[G_5]=({\rm length})^2$.

The metric $g_{\mu\nu}$ is assumed to describe a static and spherically symmetric spacetime, while the scalar field $\phi$ is taken to depend linearly on time in addition to its radial dependence \cite{Babichev:2013cya}.
This is the most general expression that one can assume for the scalar field without ruining the staticity of spacetime. {\footnote{One notable exception to this rule is flat spacetime in Fab 4 \cite{Charmousis:2011bf} theory which allows for a scalar with a $t^2$-dependence \cite{Charmousis:2014mia}. }}
Therefore, the field-content ansatz of the theory is of the form 
\begin{align}
&\mathrm{d}s^2={}-h\left(r\right)\mathrm{d}t^2+\frac{\mathrm{d}r^2}{f\left(r\right)}+r^2\mathrm{d}\Omega^2, \label{eq:metric}\\
&\phi ={} qt+\psi\left(r\right).\label{scalar}
\end{align}
For the above line element, the defining relation of the scalar field's kinetic term takes the form
\begin{equation}
\label{eq:X-inhomo}
2X = \frac{q^2}{h} - \psi'^2 f.
\end{equation}
Here and throughout this article, a prime will denote derivation with respect to the radial coordinate $r$.
To express the field equations in a concise and compact form, it is advantageous to make use of the auxiliary functions $Z(X)$ and $Y(X)$, which encapsulate specific combinations of the model functions, namely
\begin{align}
Z\left(X\right)={}&-G_4+2XG_{4X},\\[1mm]
Y\left(X\right)={}&-XG_{5X}.\end{align}
Using the above equations, we may define the functionals
\begin{align}
\mathcal{A}={}& Z_X-\frac{X}{2rf\psi'}\left(r^2G_{3X}+G_{5X}\right)+\frac{q^2}{4r h \psi'}G_{5X}+\frac{f\psi'}{4rX}(Y+2Y_X X) ,\\
\mathcal{B} ={}&Z+\frac{f\psi'}{r}Y,\end{align}
and bring the independent field equations in the following simplified form
\begin{align}
&2X'\mathcal{A} = \left(\frac{h'}{h}-\frac{f'}{f}\right)\mathcal{B}, \label{eq1}\\[2mm]
&\frac{2rf}{h}h'\mathcal{A} = r^2G_{2X}+2G_{4X}\bigg(1-\frac{q^2 f}{2hX}\bigg)-2rf\psi'G_{3X}-2fZ_X\bigg(1-\frac{q^2}{2X h}\bigg),\label{eq2}\\[2mm]
&\frac{2rf}{h}h'\mathcal{B}=-r^2G_2-2G_4\bigg(1-\frac{q^2 f}{2hX}\bigg)-2fZ\bigg(1-\frac{q^2 }{2hX}\bigg)\label{eq3}.
\end{align}


\section{Minkowski Vacuum}\label{mink}

Before examining the vacuum conditions, it is useful to express eqs. (\ref{eq1}-\ref{eq3}) in a more convenient form that simplifies calculations in the case of non-homogeneous metrics, i.e., $f\neq h$. At first, we re-express the spatial component of the metric tensor in terms of the redshift function of the compact object. We introduce  the inhomogeneity factor, $K(r)$, via 
\begin{equation}
    f(r)=h(r)[K(r)]^2.
\end{equation}
From the definition of $X$, namely eq.\,\eqref{eq:X-inhomo}, one can verify that 
\begin{equation}
    \label{eq:fpsi}
    f\psi'=\delta \sqrt{q^2-2hX}\,K, \hspace{1em} \delta=\pm 1\,.
\end{equation}
Upon utilizing the above equation, the system of differential equations takes the form 
\begin{align}
    &2(q^2-2hX)\frac{\big(\sqrt{X}YK^2\big)'}{\sqrt{X}}-X'\left[2r^2XG_{3X}+2Y \left(\frac{q^2K^2}{2X}-1\right)\right]=-4\delta r \sqrt{q^2-2hX}(Z K)'\,,\label{eqq1}\\[2mm]
    &(q^2-2hX)(KZ)'+X\left[r^2 \left(\frac{G_2}{K}\right)'-2\left(\frac{G_4}{K}\right)'\left(\frac{q^2K^2}{2X}-1\right)\right]=2\delta r X G_{3X}X'\sqrt{q^2-2hX}\,,\label{eqq2}\\[2mm]
    & KZ (q^2-2hX)-2 r X KZ h'-r^2\frac{X G_2}{K}+\frac{G_4}{K}(q^2K^2-2X)=2\delta X Y K^2 h'\sqrt{q^2-2hX}\,.\label{eqq3}
\end{align}
Note here that Eq. (\ref{eqq1}) follows from (\ref{eq1}) upon substitution of \eqref{eq:fpsi}; the same relation holds between Eqs. (\ref{eqq3}) and (\ref{eq3}). To derive Eq. (\ref{eqq2}), one needs to multiply (\ref{eq2}) by $2X'$ and (\ref{eq3}) by $(h'/h - f'/f)$, and then take their difference. By doing so, the l.h.s. of the resulting equation vanishes identically due to \eqref{eq1}, while the r.h.s corresponds to Eq. \eqref{eqq2}.

The vacuum conditions for flat spacetime can be derived from the above system of PDEs by substituting $h(r)=K(r)=1$.
Assuming that all model functions are nonzero and starting from bottom to top and utilizing at each step the result of the preceding equation, it is straightforward to determine that
\begin{tcolorbox}[enhanced,
top=5pt, bottom=5pt, left=10pt, right=10pt,
interior hidden]
    \vspace{-1em}
    \begin{align}
    \label{eq:vac-flat1}
    &r=\sqrt{\frac{2G_{4X}}{G_2}(q^2-2X)}\,,\\[2mm]
    \label{eq:vac-flat2}
    &G_3=\delta\sqrt{2G_2G_{4X}}\,,\\[2mm]
    \label{eq:vac-flat3}
    &Y\equiv-XG_{5X}=g_5 - 2\delta XG_{4X}\sqrt{\frac{2G_{4X}}{G_2}}\,,
    \end{align}
\end{tcolorbox}
\noindent where $g_5$ is a coupling constant with the same dimensions as the model function $G_5$.
Equations~(\ref{eq:vac-flat2}) and (\ref{eq:vac-flat3}) indicate that, for Minkowski vacuum, the model functions $G_3$ and $G_5$ (through $Y$) depend entirely on the choice of $G_2$ and $G_4$. 
Once a specific form for $G_2$ and $G_4$ is selected, the vacuum conditions fix the form of the remaining two functions.
Furthermore, based on the specific form of $G_2$ and $G_4$, Eq. \eqref{eq:vac-flat1} provides the functional form of $X$ in terms of the radial coordinate $r$
(the case $2X=\rm{const.}$ is separately treated in a forthcoming section).
In this sense, and since $X$ and $\phi$ acquire nontrivial profiles even in Minkowski, the solution is of stealth type. 
We adopt here the term \textit{stealth} by analogy with the corresponding stealth black-hole solutions, which correspond to GR metrics, while the scalar field remains nontrivial.
However, when $q=0$, one expects $X$ to either vanish or smoothly approach a constant value, and therefore the stealth vacuum becomes purely GR vacuum.

It is important to clarify at this point that a Horndeski theory respecting the conditions \eqref{eq:vac-flat2} and \eqref{eq:vac-flat3} does not necessarily admit only a nontrivial Minkowski vacuum.
In principle, it may support a broader class of static solutions, of lesser symmetry, including even non-homogeneous configurations (eg. black holes, wormholes, naked singularities).
However, such solutions will not satisfy the condition \eqref{eq:vac-flat1}. 
Nevertheless, at the Minkowski limit of these solutions, the kinetic term, $X$, reduces to that given by \eqref{eq:vac-flat1}, ensuring that all three conditions are simultaneously satisfied. In other words, \eqref{eq:vac-flat1} characterizes the scalar of our maximally symmetric vacuum. 
Furthermore, if a given theory admits a black-hole solution (whether homogeneous or not), then the Minkowski vacuum obtained here can be interpreted as the limiting case in which the black hole mass approaches zero.
In this limit, the solution continuously approaches the vacuum configuration $h=K=1$, while the scalar field remains nontrivial.
Below, we present two illustrative examples of asymptotically flat black-hole solutions with a non-GR Minkowski vacuum. 
The first example corresponds to a homogeneous solution, whereas the second involves a non-homogeneous one.



\subsection{Extension of the regularized EGB black hole}
\label{Sec:Min}

The regularized four-dimensional Einstein-Gauss-Bonnet (EGB) theory is obtained as a singular limit of Lovelock theory (in the present context see \cite{Charmousis:2014mia}). After the initial limit proposed by Glavan and Lin \cite{Glavan:2019inb} the careful construction of the scalar-tensor theory was undertaken using different methods in \cite{Lu:2020iav, Fernandes:2020nbq,Hennigar:2020lsl, Fernandes:2021dsb} and subsequently homogeneous black hole solutions were also found. The theory involves all Horndeski functionals and reads
\begin{equation}
    \label{eq:LP-G24}
    G_2=8\alpha X^2 \quad \text{and} \quad G_4=1+4\alpha X\,,
\end{equation}
while one can determine the functional form for $G_3$ and $G_5$ using Eqs. \eqref{eq:vac-flat2} and \eqref{eq:vac-flat3}, respectively. 
By doing so, and choosing $\delta=-1$ and $g_5=-4\alpha$, we are led to
\begin{equation}
    \label{eq:LP-G35}
    G_3=-8\alpha X \quad \text{and} \quad G_5=-4\alpha\ln |X|\,.
\end{equation}
Here, $\alpha$ is a positive coupling constant of dimensionality $({\rm length})^2$.
Given the shift invariance of the scalar field, one can extend the scalar anzatz to \eqref{scalar} with $q\neq 0$ and find the following solution to the field equations \cite{Charmousis:2021npl},
\begin{equation}
   h(r)=f(r)=1+\frac{r^2}{2 \alpha}\left(1-\sqrt{1+\frac{8\alpha M}{r^3}}\right) \quad {\rm and} \quad 
   X(r)=\frac{2\sqrt{f(r)+q^2 r^2}-1-f(r)}{2 r^2}\,,
   \label{eq:LP-hX}
\end{equation}
The above line element describes a static and asymptotically flat black hole of mass $M$.\,\footnote{We have intentionally ignored the asymptotically de Sitter branch since it is not relevant here.}.
The spacetime geometry is identical to the $q=0$ solution  presented in \cite{Lu:2020iav, Fernandes:2020nbq, Hennigar:2020lsl,Fernandes:2021dsb}. In fact when $M=q=0$ the scalar field is constant and the Minkowski vacuum is identical to the GR vacuum.
When $q\neq 0$ on the other hand, the modification of the scalar is not trivial as it renders the solution regular at the horizon and well within (depending on the magnitude of $q$).
The second important point is that for $q\neq 0$ and setting $M=0$, we recover flat Minkowski geometry, however, the kinetic term of the scalar field remains nontrivial and is given by the expression
\begin{equation}
    \label{eq:LP-X0}
    X(r)\big|_{M=0}=\frac{q^2}{1+\sqrt{1+q^2 r^2}}\,.
\end{equation}
This is therefore a stealth Minkowski vacuum as long as $q\neq 0$.
Furthermore, notice that for $M=0$, $X$ is regular for all $r$, while one can also verify that Eq. \eqref{eq:vac-flat1} is satisfied.
This is the only explicit nontrivial black-hole solution featuring all Horndeski functionals and we will refer to it as the qEGB solution \eqref{eq:LP-hX}. The black-hole geometry is a singular limit of the celebrated Boulware-Deser solution \cite{Boulware:1985wk} of Lovelock theory. In this sense, and since the integration constant $q$ does not appear in the geometry it can be loosely referred to as a stealth Lovelock black hole. We will see in later sections how disformal transformations of this solution will give $q,M$ dependent spacetime metrics.


\subsection{Non-homogeneous black-hole solution}

As a second example, we consider the family of theories with model functions
\begin{align}
\label{eq:vac2-G24}
G_2 = \frac{4 \alpha n}{\eta^2} X^n \qquad\text{and}\qquad G_4 = 1 + \alpha X^n,
\end{align}
where $\alpha$ and $\eta$ are coupling constants with dimensions $({\rm length})^{2n}$ and $({\rm length})$, respectively.
Assuming also that $\alpha$, $\eta$, and $n$ are positive-valued and employing Eqs. \eqref{eq:vac-flat2}, \eqref{eq:vac-flat3}, we can readily obtain
\begin{align}
    \label{eq:vac2-G35}
    G_3 = 2\sqrt{2}\, \frac{\alpha n\delta }{\eta}  X^{n - \frac{1}{2}} \qquad\text{and}\qquad G_5 =\delta\frac{\sqrt{2}\,n \alpha \eta}{n-1/2}      X^{n - \frac{1}{2}}, \qquad n\neq1/2
\end{align}
In the above, $g_5$ was assumed to be zero, while we remind the reader that $\delta=\pm 1$.
The special case $n=1/2$ gives trivial $G_3$ and will be discussed in the subsequent section.
Finally, from Eq.~(\ref{eq:vac-flat1}) we find the expression for \( X \) corresponding to the Minkowski vacuum, which is given by 
\begin{equation}
    \label{eq:vac-flat-X}
    X = \frac{q^2}{2}\frac{1}{1+r^2/\eta^2}.
\end{equation}
As in the previous case, for $q = 0$ the scalar field and its kinetic term $X$ vanish identically, and the Horndeski Minkowski vacuum reduces to the GR Minkowski vacuum.  Interestingly enough, the functional form of $X$ is the one encountered in the beyond Horndeski family \cite{Bakopoulos:2023fmv, Bakopoulos:2023sdm} with $G_2,G_4$ given by \eqref{eq:vac2-G24} and trivial $G_3,G_5$. Therefore one can think of this class as its Horndeski extension to nontrivial $G_3, G_5$ \eqref{eq:vac2-G35} with an important difference: Black holes of the former theory acquired the same $X$ as the Minkowski vacuum, however here, the latter theory will admit a mass dependent $X$ functional as we now elucidate.

To investigate general solutions of the theory \eqref{eq:vac2-G24} and \eqref{eq:vac2-G35}, beyond the vacuum, let us focus on the linear case $n = 1$. The non-homogeneous system (\ref{eqq1}-\ref{eqq3}) cannot be solved analytically for arbitrary $h$, $K$, and $X$ and we employ numerical methods.
For  $n = 1$, we can verify that the general solution at large $r$ takes the following form:
\begin{align}
    h(r) & \rightarrow 1 - \frac{2M}{r} + \frac{\alpha \eta^2 M q^2}{r^3} - \frac{3 \alpha \eta^4 M q^2 (\alpha q^2 + 1)}{5 r^5} + \frac{\alpha \eta^4 M^2 q^2}{r^6} + \mathcal{O}\left(\frac{1}{r^7}\right),\label{inf1}\\[2mm]
    K(r) &\rightarrow 1 - \frac{\alpha \eta^4 M q^2}{5 r^5} + \frac{\alpha \eta^6 M q^2 (\alpha q^2 + 2)}{7 r^7} + \frac{\alpha \eta^6 M^2 q^2}{16 r^8} + \mathcal{O}\left(\frac{1}{r^9}\right),\label{inf2}\\[2mm]
    X(r) &\rightarrow \frac{q^2 \eta^2}{2r^2} - \frac{\eta^4 q^2}{2 r^4} + \frac{\eta^6 q^2}{2 r^6} - \frac{\eta^6 q^2 (\eta^2 + 3M^2)}{2 r^8} + \mathcal{O}\left(\frac{1}{r^9}\right),\label{inf3}
\end{align}
where $M$ is the ADM mass. From the asymptotic expansion, we observe that the solution is indeed non-homogeneous since $K\neq1$. However, at the asymptotic regime, the solution closely resembles the Schwarzschild solution, as the deviation from homogeneity is extremely mild.
Notice that the first correction in $K$ appears at order $\mathcal{O}(r^{-5})$. 
Moreover, since $q$ appears in the metric as an independent parameter, the solutions are characterized by a primary charge, but contrary to the solutions presented in \cite{Bakopoulos:2023fmv, Bakopoulos:2023sdm}, here, $q$ does not appear independently in the general solution, but always in the combination $qM$. Note also that $X$ explicitly depends on the ADM mass $M$ unlike the solutions found in \cite{Bakopoulos:2023fmv, Bakopoulos:2023sdm}.

\begin{figure}[t]
    \centering
    \begin{subfigure}[b]{0.476\textwidth}
    \includegraphics[width=1\textwidth]{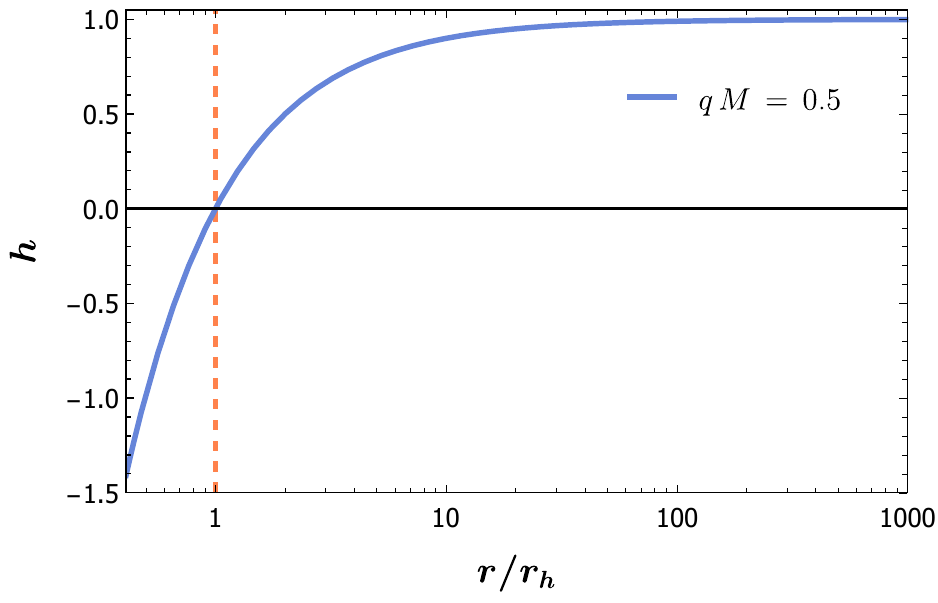}
    \caption{\hspace*{-2.4em}}
    \label{subf:h1}
    \end{subfigure}
    \hfill
    \begin{subfigure}[b]{0.48\textwidth}
    \includegraphics[width=1\textwidth]{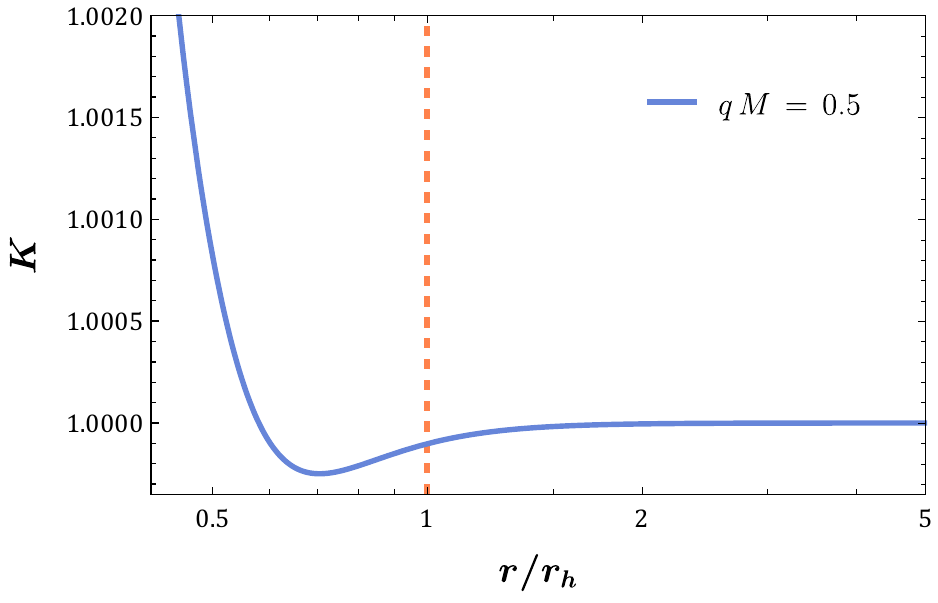}
    \caption{\hspace*{-4em}}
    \label{subf:k1}
    \end{subfigure}
    \vspace*{-0.5em}
    \caption{(a) The metric function $h$, and (b) the inhomogeneity factor $K$ for $q M=0.5$, $\eta/M=1$, $\alpha/M^2=0.1$. The horizontal axis is logarithmic in both graphs.}
    \label{fig:hk}
\end{figure}

Finding an asymptotic solution does not guarantee the existence of a global solution. To check whether a a global solution exists for all $r\in(0,+\infty)$, we solve the system (\ref{eqq1}-\ref{eqq3}) numerically, using the asymptotic expansions (\ref{inf1}-\ref{inf3}) as initial conditions. Note that this is a differential-algebraic system, since Eqs. (\ref{eqq1}-\ref{eqq2}) are algebraic in $h$.
In Fig. \ref{fig:hk}, we present a solution for $q M = 0.5$. The figure shows the behavior of the metric function $h$ and the inhomogeneity factor $K$. We observe that the solution is asymptotically flat, as expected. 
It possesses a horizon at a finite distance $r_h$ and a singularity at the origin. From the form of the inhomogeneity factor, it is evident that the solution deviates only slightly from the homogeneous case, something already apparent from the asymptotic expansion. 
Notably, for larger values of $n$, the corresponding solutions are even more homogeneous. 
For example, when $n = 2$, the first correction in $K$ appears at order $r^{-7}$, as opposed to $r^{-5}$ in the $n = 1$ case. Nevertheless, there will always be a nonzero deviation from the homogeneous case, despite the fact that $K$ tends to approach unity as $n$ increases.


\subsection{Special cases}

The conditions \eqref{eq:vac-flat2} and \eqref{eq:vac-flat3} were derived under the assumption that all Horndeski functions are non-vanishing. 
However, before we conclude this section, it is important to discuss several special cases in which one or two of the model functions are absent and thus the conditions for Minkowski vacuum are not necessarily given by \eqref{eq:vac-flat2} and \eqref{eq:vac-flat3}. 
These cases not only lead to simplifications but also highlight interesting structures and known results within the theory.
We begin with the case where $Y = 0$, which corresponds to setting $G_{5X} = 0$. 
In this case, the conditions \eqref{eq:vac-flat3} and \eqref{eq:vac-flat2} reduce to 
\begin{tcolorbox}[enhanced,
top=5pt, bottom=5pt, left=10pt, right=10pt,
interior hidden]
    \vspace{-1.5em}
\begin{gather}
    \text{\underline{$G_{5X}=0$ subclass}}\nonumber\\[3mm]
    \label{eq:flat-no-G5}
    G_2=8g_2^2 X^2 G_{4X}^3 \quad\text{and}\quad G_{3X}=4\delta |g_2 X| G^2_{4X},
\end{gather}
\end{tcolorbox}
\noindent respectively. The coupling constant $g_2=1/g_5$ is assumed to be positive with the same dimensions as the model function $G_2$. 
This structure reveals that the theory is essentially characterized by a single model function, with the remaining functions determined by it.
Next, consider the case where $G_{3X} = 0$ while $G_{5X}$ remains non-zero. The conditions in this case become
\begin{tcolorbox}[enhanced,
top=5pt, bottom=5pt, left=10pt, right=10pt,
interior hidden]
    \vspace{-1.5em}
\begin{gather}
    \text{\underline{$G_{3X}=0$ subclass}}\nonumber\\[3mm]
    G_2=\frac{\xi}{G_{4X}}\quad \text{and}\quad G_{5X}=-\frac{g_5}{X}- 2\delta\sqrt{\frac{2}{\xi}}G^2_{4X},
\end{gather}
\end{tcolorbox}
\noindent where $\xi$ is a dimensionless coupling constant and $g_5$ is a coupling constant with the same dimensions as $G_5$. 
Once again, the theory is characterized by a single free function, with the remaining functions constrained accordingly. This case for $n=1/2$ includes the  singular case encountered in the previous section \eqref{eq:vac2-G35} and we shall briefly discuss it in a moment.
Finally, we turn to the case where $G_{3X} =G_{5X} = 0$, corresponding to a $\bold{Z}_2$-symmetric theory. The vacuum conditions now simplify dramatically 
\begin{tcolorbox}[enhanced,
top=5pt, bottom=5pt, left=10pt, right=10pt,
interior hidden]
    \vspace{-1.5em}
\begin{gather}
    \text{\underline{$G_{3X}=G_{5X}=0$ subclass}}\nonumber\\[3mm]
     G_2=\frac{\xi}{\alpha} \sqrt{X} \quad \text{and}\quad  G_4=1+2\alpha \sqrt{X},
\end{gather}
\end{tcolorbox}
\noindent where $\xi$ is a dimensionless coupling constant, while $\alpha$ carries units of (length). 
This case is particularly noteworthy because it corresponds to a unique and well-known theory that admits the stealth Schwarzschild solution as an exact solution \cite{Bakopoulos:2023fmv}. This is the only case where the scalar field remains the same for $M\neq 0$. 
In fact, for $G_4=1+2\alpha \sqrt{X}$, the $G_{3X}=0$ subclass leads to the same $G_2,G_4$ but with a nontrivial $G_5\sim \ln{|X|}$.  
When the $G_5$ term is added to the theory the black-hole solution ceases to be stealth and homogeneous. An asymptotic expansion for large $r$ gives,
\begin{align}
    h(r) &\rightarrow 1 - \frac{2M}{r} + \frac{\sqrt{2}\alpha\eta  M q }{r^2}  +\mathcal{O}\left(\frac{1}{r^3}\right),\label{inf11}\\[2mm]
    K(r) &\rightarrow 1 - \frac{\alpha \eta^3 M q}{4 \sqrt{2}\, r^4} +  + \mathcal{O}\left(\frac{1}{r^5}\right),\label{inf21}\\[2mm]
    X(r) &\rightarrow \frac{q^2 \eta^2}{2r^2} - \frac{\eta^4 q^2}{2 r^4} + \frac{\eta^5 q^2 (2 \sqrt{2}\eta-\alpha M q)}{4\sqrt{2}\, r^6} + \mathcal{O}\left(\frac{1}{r^9}\right).\label{inf31}
\end{align}
In all cases with Minkowski vacuum, the scalar field is given by Eq. \eqref{eq:vac-flat1}. 

Finally, since the class of theories considered here lead to nontrivial profiles for the scalar field, one can exploit this feature to reformulate the system in terms of the scalar invariant $X$. Specifically, by using the parametrization $R(X) = r$ and $H(X) = h$, all quantities can be expressed in terms of $X$. This reparameterization can significantly streamline both analytic and numerical treatments of the solutions, as it replaces the radial dependence with the scalar field profile itself. 


\section{Field-theoretic constraints for homogeneous solutions}\label{yo}

In this section, we investigate the conditions under which a model in shift-symmetric Horndeski gravity admits static and spherically symmetric homogeneous solutions, namely $h(r)=f(r)$ in \eqref{eq:metric}. The equivalent analysis for the case $q=0$ and beyond Horndeski theories was undertaken in \cite{Bakopoulos:2022csr}.
It turns out that our analysis here is completely general and the constraints obtained are not subject to a particular ansatz for the scalar field or the metric. The constraints stem from imposing internal consistency of the field equations under the assumption of homogeneity.
In particular, for static and spherically symmetric spacetimes, the field equations reduce to a set of three independent differential equations (\ref{eq1}-\ref{eq3}). 
However, when homogeneity is imposed, at the level of the metric, the number of independent metric components (and hence the number of unknown functions) is reduced to two: the function $h(r)$ and the scalar invariant $X(r)$. 
Consequently, the system becomes overdetermined: three equations must be satisfied by only two functions. 
This overdetermination implies that two of the field equations must be algebraically compatible with one another, otherwise no consistent solution can exist within the assumed class.
The requirement of compatibility between two of the three field equations imposes a set of nontrivial algebraic relations among the Horndeski model functions. 
These relations, which we derive explicitly in this section, serve as necessary conditions for the existence of homogeneous solutions, independently of the characteristics of the solutions.
In this sense, the compatibility conditions act as a filter and restrict the allowed phase space of theories before any further analysis.
Therefore, the study of these constraints is a critical first step in the classification of viable Horndeski models that admit homogeneous solutions.

Since we are interested in the homogeneous case ($f=h$), it is more convenient to work with the system of Eqs. (\ref{eq1} -\ref{eq3}). 
Specifically, Eqs. (\ref{eq1}) and (\ref{eq2}), take the form
\begin{align}
   &  (h\psi')^2\, - 4r\frac{X Z_X }{Q} (h\psi')  +\frac{2X^2}{Q}\left[r^2G_{3X}+G_{5X}\left(1-\frac{q^2}{2X}\right)\right]=0,\label{feeq}\\[2mm]
   &  (h \psi')^2-2r\frac{X G_{3X}}{Z_X} (h\psi') +\frac{X}{Z_X}\left[r^2 G_{2X}+2G_{4X}\left(1-\frac{q^2 }{2X}\right)\right]=0,\label{seeq}
\end{align}
where we have defined $ Q(X)\equiv-(Y+2X Y_X)$ and also made use of \eqref{eq:X-inhomo}. By imposing compatibility of these two equations, we get the following conditions for the model functions of the theory
\begin{gather}
    G_{3X}=\frac{2 Z_X^2}{Q},\hspace{1.5em}
    G_{2X}=\frac{4 X Z_X^3}{Q^2},\hspace{1.5em}
    G_{4X}=\frac{X Z_X G_{5X}}{Q}\,.
\end{gather}
Substituting the expressions for $Q(X)$ and $Y(X)$, the last equation yields $G_{4X}=\xi Y=-\xi X G_{5X}$, where $\xi$ serves as a dimensionless coupling constant. 
Utilizing the last relation, one can verify that $Q=-Z_X/\xi$, which by its turn leads to the following simplified conditions for the model functions of the theory
\begin{tcolorbox}[enhanced,
top=5pt, bottom=5pt, left=10pt, right=10pt,
interior hidden]
    \vspace{-1em}
    \begin{align}
    \label{eq:hom-conds}
    G_{2X}=4\xi^2 X Z_X\,, \hspace{1.5em}
    G_{3X}=-2\xi Z_X\,, \hspace{1.5em}
    G_{4X}=-\xi X G_{5X}\,.
    \end{align}
\end{tcolorbox}
\noindent From the above relations, and the fact that $Z$ depends solely on $G_4$, it follows that a Horndeski theory admitting homogeneous solutions contains just a single free model function, with conditions \eqref{eq:hom-conds} determining the remaining ones.
We can now solve either \eqref{feeq} or \eqref{seeq} in terms of $(h\psi')$.
By doing so, and employing also the compatibility conditions \eqref{eq:hom-conds}, it is straightforward to find
\begin{equation}
    \label{hpsi}
    h\psi'= -2 \xi r X +\delta \sqrt{q^2-2X}\sqrt{\frac{G_{4X}}{Z_X}}\,, \hspace{1em} \delta=\pm 1\,.
\end{equation}
From the above relation and the definition of the kinetic term \eqref{eq:X-inhomo}, one can readily determine the metric function $h(r)$, which for homogeneous solutions takes the form
\begin{tcolorbox}[enhanced,
top=5pt, bottom=5pt, left=10pt, right=10pt,
interior hidden]
    \vspace{-1em}
\begin{equation}
    \label{metric}
    h(r)\equiv \frac{q^2-(h\psi')^2}{2X}=1-\frac{q^2-2X}{2X}\frac{G_{4X}-Z_X}{Z_X}+2\delta \xi r\sqrt{q^2-2X}\sqrt{\frac{G_{4X}}{Z_X}} -2\xi^2 r^2 X.
\end{equation}
\end{tcolorbox}

The above result implies that the metric component and the kinetic term of the scalar field $X$ are completely intertwined when the functionals $G_3$ and $G_5$ of the $\bold{Z}_2$-breaking Horndeski gravity are not trivial. 
An important consequence of the latter is that the final equation of motion \eqref{eq3} can be integrated with respect to $X$, thereby reducing it to an algebraic equation in $r$.
Making use of the compatibility conditions \eqref{eq:hom-conds} and the fact that $W(X)X'\equiv  \partial_r \int \diff X W(X)$, one can verify that \eqref{eq3} can be brought to the form 
\begin{tcolorbox}[enhanced,
top=5pt, bottom=5pt, left=10pt, right=10pt,
interior hidden]
    \vspace{-1em}
\begin{equation}
 \label{eq:hom-int}
    2\xi^3 r^3\int \diff X G_{4}-2 \xi^2 r^2G_4 \mathcal{G}+\xi r \mathcal{G}^2 G_4\left(\frac{G_{4X}}{G_4}-\frac{G_{4XX}}{G_{4X}}\right) +\int \diff X G_{4X} \mathcal{G} \left[\mathcal{G}^2\frac{G_{4XX}}{G_{4X}}\right]_X={\rm const.},
\end{equation}
\end{tcolorbox}
\noindent where 
\begin{equation}
    \mathcal{G}\equiv \delta \sqrt{(q^2-2X)\frac{G_{4X}}{Z_X}}.
\end{equation}
We emphasize that the last equation applies to all Horndeski gravities that respect the compatibility conditions \eqref{eq:hom-conds}.
This expression represents the most compact and analytically manageable form in which the equations of motion can be recast for the shift-symmetric class of Horndeski theories. 
By expressing all dynamical quantities in terms of the scalar kinetic term $X$, the system of coupled differential equations can be reduced to a single nonlinear integral equation \eqref{eq:hom-int}.
This equation encapsulates the full dynamical content of the theory and provides a clear pathway for exploring the solution space. 
Although this reduction significantly simplifies the structure of the system, it does not guarantee the existence of analytic solutions. 
The resulting equation is, in general, highly nonlinear. 
While equation \eqref{eq:hom-int} admits solutions of the inverted polynomial form $r(X)$, yielding the roots of the complete homogeneous solution space in $\bold{Z}_2$-breaking Horndeski gravity, the cubic nature of the equation implies that $r(X)$ is not, in general, bijective. 
As a result, the geometry cannot be analytically represented in Schwarzschild coordinates.

Finally, we stress that in order for a given Horndeski theory to admit a Minkowski vacuum solution that is also homogeneous, it must satisfy both sets of conditions simultaneously: the vacuum conditions derived in the previous section and the compatibility conditions established here.
This dual requirement places strong constraints on the allowed functional forms of the theory, significantly narrowing the class of admissible models. 
The simultaneous satisfaction of both conditions is not trivial, as each set independently imposes algebraic relations among the Horndeski functionals, and their intersection is highly restrictive. 
Remarkably, the only theory satisfying both sets of conditions is the regularised EGB theory \cite{Lu:2020iav,Fernandes:2020nbq,Hennigar:2020lsl,Fernandes:2021dsb}, which was presented in the preceding section as well.


\section{De Sitter Vacuum}\label{yo1}

We now turn our attention to the existence of de Sitter vacuum solutions within the shift-symmetric Horndeski theory. This analysis serves as a natural extension of the previous sections, providing insight into scalar-tensor configurations with a cosmological constant. 
To this end, we consider a fixed background geometry characterized by the metric functions
\begin{equation}
h(r) = 1 - \Lambda r^2,\qquad K(r) = 1,
\end{equation}
which describe a static patch of de Sitter spacetime with a positive cosmological constant $\Lambda$. 
We restrict ourselves to the subclass of theories with $Y = 0$, which corresponds to the choice $G_{5X} = 0$. 
This class of theories has been extensively studied in the context of cosmic acceleration, as it can provide, in certain cases, viable dark energy models. In particular if scalar-vector-tensor (SVT) theories are considered the multi-messenger constraints (see for example \cite{Sakstein:2017xjx, Ezquiaga:2017ekz,Creminelli:2017sry} and also \cite{deRham:2018red}) can be strongly suppressed. This is achieved by considering non-trivial couplings between the dark energy scalar and the electromagnetic sector which maintains $U(1)$ symmetry but has variable speed of light at strongly curved backgrounds (see for example \cite{Mironov:2024idn, Mironov:2024wbx, Babichev:2024kfo, Mironov:2025dzz}).
Our goal here is to identify the conditions within our theory, under which the model functions of the theory support the de Sitter spacetime as a solution.
Solving the system \eqref{eqq1}-\eqref{eqq3}, under the above assumptions, the equations reduce to the following algebraic system:
\begin{tcolorbox}[enhanced,
top=5pt, bottom=5pt, left=10pt, right=10pt,
interior hidden]
    \vspace{-1em}
\begin{align}
&r =\sqrt{\frac{2(q^2-2X)}{g_2^2 X^2 G_{4X}^2}}, \label{eq:ds1} \\[2mm]
&G_{3X} =\delta\frac{Z_X}{X}\sqrt{2g_2^2 X^2 G_{4X}^2+8\Lambda X}
, \label{eq:ds2}\\[2mm]
&G_2 = 6 \Lambda Z + g_2^2 X^2 G_{4X}^3, \label{eq:ds3}
\end{align}
\end{tcolorbox}
\noindent where $g_2$ is a coupling constant with the same dimensions as $G_2$, and $\delta = \pm 1$. 
Once again, the structure of the system is purely algebraic and depends only on the scalar quantity $X$.

Given that $G_{5X}=0$, the above system dictates that, for de Sitter vacuum, the theory contains only one independent model function.
Once this function is specified, the entire theory is determined via Eqs. \eqref{eq:ds1}-\eqref{eq:ds3}, together with the expression of the kinetic term $X$ in terms of the radial coordinate $r$.
In direct analogy with the Minkowski vacuum, Eqs. \eqref{eq:ds1}-\eqref{eq:ds3} define a subclass of shift-symmetric Horndeski theories that admit the de Sitter geometry as a vacuum solution.
Notice that for $\Lambda\to 0$, Eqs. \eqref{eq:ds1}-\eqref{eq:ds3} reduce to the relations \eqref{eq:vac-flat1} and \eqref{eq:flat-no-G5}.
Consequently, it follows that the inclusion of de Sitter configurations increases the complexity of the admissible theory space.


\section{Solutions with constant kinetic term}\label{yo2}

In this section, we follow a different approach and we consider theories with constant kinetic term, namely $X'=0$.
In the past, such theories led to numerous stealth solutions starting with those found in \cite{Babichev:2013cya, Kobayashi:2014eva}. 
Having written the general equations in a very compact form, the algorithm to follow is quite straightforward. 
We start from \eqref{eq1} where we have that either $h=f$ or $\mathcal{B}=0$. 
The latter possibility leads to problematic or strongly coupled cases so we choose the former to proceed. 
Assuming that $2X=\tilde{q}^2$, with $\tilde{q}\neq q$, the definition \eqref{eq:X-inhomo} leads us to
\begin{equation}
    \label{eq:F}
    f \psi'=\delta \sqrt{q^2-\tilde{q}^2 f}=\delta \tilde{q} F(r),\qquad \delta=\pm 1\,,
\end{equation}
where the last equality defines the convenient function $F(r)\equiv \sqrt{q^2/\tilde{q}^2-f(r)}$\,.
Without loss of generality, we take both $q$ and $\tilde{q}$ to be positive to avoid expressions with absolute values.
It is also important to mention that the assumption $2X=q^2$, with $\tilde{q}=q$ is not restrictive. 
One can verify that even in this scenario, the solutions resulting from the field equations would be of the same form, albeit with different constraints on the model functions. 
Notably, in certain cases the condition $\tilde{q}=q$ is imposed to us for consistency of the field equations.
By employing \eqref{eq:F} in the remaining field equations \eqref{eq2} and \eqref{eq3}, one can readily show that they can be integrated once, giving in turn
\begin{tcolorbox}[enhanced,
top=5pt, bottom=5pt, left=10pt, right=10pt,
interior hidden]
    \vspace{-1em}
\begin{gather}
    \label{eq:X-con-1}
   \frac{\delta}{3\tilde{q}r}(Y+ \tilde{q} ^2 Y_X) F^3+ Z_X F^2-\frac{\delta \tilde{q}}{2} \left[ r G_{3X}+\bigg(1-\frac{q^2}{\tilde{q}^2}\bigg) \frac{G_{5X}}{r}\right]F+\frac{G_{2X}}{6}r^2+G_{4X}\bigg(1-\frac{q^2}{\tilde{q}^2}\bigg) -\frac{C_1}{2r}=0,\\[3mm]
  \label{eq:X-con-2}
  \frac{2\delta}{3} \frac{\tilde{q}}{r}YF^3 + Z F^2- G_{4}\bigg(1-\frac{q^2}{\tilde{q}^2}\bigg) -\frac{G_2}{6} r^2 -\frac{C_2}{2r}=0\,,
\end{gather}
\end{tcolorbox}
\noindent with $C_1,\,C_2$ integration constants. 
Both equations are algebraic with respect to the variable $F$.
Since the system of equations involves third-order polynomials, it cannot be solved straightforwardly.
Therefore, in what follows we present a few indicative examples of vacuum solutions.

\subsection*{$\bold{Z}_2$-symmetric theories: $G_3=G_5=0$} 

The first set of solutions is for the $\bold{Z}_2$-symmetric theories with $G_3=G_5=Y=0$.
In this case, for $\tilde{q}=q$ and $ZG_2=\lambda$, the above equations host as solution the Schwarzschild-de Sitter geometry
\begin{equation}
    f(r)=1-\frac{2M}{r}-\frac{\lambda}{6Z^2}r^2\,,
\end{equation}
with $\lambda$ an arbitrary coupling constant.
Solutions of this type have been widely studied \cite{Babichev:2013cya, Kobayashi:2014eva, Charmousis:2015aya, Minamitsuji:2018vuw, Takahashi:2020hso}.


\subsection*{$\bold{Z}_2$-breaking theories with $G_3$}

In addition to the previous case, we now allow for a $G_3$ term while setting $\tilde{q}=q$ and $G_5=Y=0$. 
It is common lore that the presence of $G_3$ alone makes explicit solutions for black holes non-existent to date. 
Numerical solutions for static and rotating solutions have been found \cite{Babichev:2016kdt, VanAelst:2019kku}. It is then interesting to note that by setting $C_1=C_2=0$, we can have de Sitter self-tuning theories \cite{Klinkhamer:2007pe, Charmousis:2011bf, Charmousis:2015aya, Martin-Moruno:2015bda, Babichev:2016kdt} with
\begin{equation}
\label{desitter}
    f(r)=1-\lambda r^2\,,
\end{equation}
and theories defined by $G_{2X}=\frac{3 X G_{3X}^2}{4 Z_X}$ and $\lambda=\frac{G_2}{6 Z}$. 
In the preceding section, we have examined how these de Sitter vacua can be found in theories with nontrivial $X$. 
This class of theories are not the ones found and analysed in great detail in \cite{Babichev:2012re} as this paper treats the case with $G_4=-Z=1$. The Babichev-Far\`ese class boils down to additionally assuming $Z_X=0$ which gives us the theory constraints $\frac{G_2}{6 Z}=\frac{2 G_{2X}^2}{9 X G_{3X}^2}=\lambda$ with a static de Sitter solution \eqref{desitter}.


\subsection*{$\bold{Z}_2$-breaking theories with $G_5$}

For the last case we set $G_{3}=0$ and also $\tilde{q}=q$. 
We then have a third order polynomial equations with respect to $F$. The two equations are compatible for theory functions such that, 
\begin{equation}
    G_2 Z=\lambda, \qquad \frac{Y+2X Y_X}{Y}=4X\frac{Z_X}{Z}\,.
\end{equation}
The resulting solution respects the following third-order equation with respect to $F$:
\begin{equation}
    \frac{2\delta q}{3r}YF^{3}+Z F^2-\frac{G_2}{6}r^2-\frac{C}{2r}=0\,.
\end{equation}
The general solution for the above equation is not trivial.
However, one can easily verify that, as in the previous cases, for $C=0$, the de Sitter geometry also constitutes a solution to the above algebraic equation, with the cosmological constant being expressed in terms of the model functions of the theory.
Another interesting scenario arises when $G_2$ is taken to be zero. 
In this case, the resulting solution is always asymptotically flat and is determined by a simplified third-order polynomial. The solution could describe black holes depending on the choice of model functions.


\section{From vacuum to compact-object solutions through disformal transformation}\label{yo3}

Having examined solutions with nontrivial vacua in the context of shift-symmetric Horndeski theory, we next investigate the characteristics of such solutions under a disformal transformation.
Do they maintain their nontrivial vacuum profile?
A disformal transformation of the metric tensor, involving a function $D=D(X)$, maps a seed Horndeski solution to a beyond-Horndeski one \cite{Zumalacarregui:2013pma, Gleyzes:2014dya, Langlois:2015cwa, Langlois:2018dxi}. For shift-symmetric theories it is only when $D$ is a constant that the image theory is still a Horndeski theory.
Consider for instance, a generic Horndeski theory with model functions $\{\bar{G}_2(\bar{X}),\, \bar{G}_3(\bar{X}),\, \bar{G}_4(\bar{X}),\, \bar{G}_5(\bar{X}) \}$, metric functions $\bar{h}(r), \bar{f}(r)$, and scalar field $\bar{\phi}=qt +\bar{\psi}(r)$. 
Here and throughout this section, barred quantities refer to the seed Horndeski theory, while unbarred quantities denote those of the target beyond-Horndeski theory.

A disformal transformation of the form
\begin{equation}
\label{disformal}
g_{\mu \nu}=\bar{g}_{\mu \nu}-D(\bar{X})\; \partial_\mu \bar \phi \,\partial_\nu \bar \phi\,,    
\end{equation}
leads to a solution in a shift-symmetric beyond-Horndeski theory, with model functions \{$G_i(\bar{X}), F_4(\bar{X}), F_5(\bar{X})$\} determined by the following relations:
\begin{align}
\label{g2g3d}
&G_2=\frac{\bar{G_2}}{(1+2 \bar{X} D)^{1/2}}\,,\qquad
G_{3X}=\bar{G}_{3\bar{X}}\frac{(1+2 \bar{X} D)^{5/2}}{1-2 \bar{X}^2 D_{\bar{X}}}\,,\\[2mm]
\label{g4g5d}
&G_4=\frac{\bar{G_4}}{(1+2 \bar{X} D)^{1/2}}\,, \qquad 
G_{5X}=\frac{\bar{G}_{5\bar{X}}(1+2 \bar{X} D)^{5/2}}{1-2 \bar{X}^2 D_{\bar{X}}}\,,\\[2mm]
\label{f4d}
&F_4=(\bar{G_4}-2\bar{X}\bar{G}_{4\bar{X}})\frac{D_{\bar{X}}(1+2 \bar{X} D)^{5/2}}{2(1-2 \bar{X}^2 D_{\bar{X}})}\,,\\[2mm]
\label{f5d}
&F_5=\bar{X}\bar{G}_{5\bar{X}}\frac{D_{\bar{X}}(1+2 \bar{X} D)^{7/2}}{6(1-2 \bar{X}^2 D_{\bar{X}})}\,.
\end{align}
Note that a legitimate disformal transformation is characterized by $1+2D(\bar{X})\bar{X}>0$ for all positive $r$.
The transformations related to the kinetic term and the disformal factor $D(X)$, are given by
\begin{gather}
\label{Xd}
X=\frac{\bar{X}}{1+2 D \bar{X}},\qquad 
\frac{d \bar{X}}{d X}=\frac{(1+2 D \bar{X})^2}{1-2 \bar{X}^2 D_{\bar{X}}}\,,\qquad
D_X=D_{\bar{X}} \frac{(1+2 \bar{X} D)^2}{1-2 \bar{X}^2 D_{\bar{X}}}\,,
\end{gather}
where in case of an invertible function $X(\bar{X})$, we can express the model function $\{G_i,\, F_4,\, F_5\}$ in terms of the kinetic term $X$.
The line element and the expression for the scalar field $\phi$, after the disformal transformation, are given by
\begin{gather}
    \label{eq:dis-metr}
    \diff s^2=-h(r)\diff \tau^2+\frac{\diff r^2}{f(r)}+r^2\diff \Omega^2\,,\qquad h=\bar{h}+D q^2\,,\qquad f=\frac{\bar{f}}{\bar{h}}\frac{h}{1+2 D \bar{X}} \,,\\[2mm]
    \phi=q\tau +\int \diff r \; \frac{\bar{h}}{h}\bar{\psi'}\,.
\label{fxd}
\end{gather}
Notice that we have made the resulting metric diagonal by rescaling time as $\tau=t+\int \diff r \frac{D q \bar{\psi'}}{\bar{h}+D q^2}$. 

The seed Horndeski theory and solution that we consider is the asymptotically flat branch of the qEGB solution \cite{Charmousis:2021npl}, which has been discussed in Sec.\,\ref{Sec:Min}. 
The Horndeski model functions are of the form \cite{Lu:2020iav,Fernandes:2020nbq,Hennigar:2020lsl,Fernandes:2021dsb}
\begin{equation}
    G_2=8\alpha \bar{X}^2,\; G_3=-8\alpha \bar{X},\; G_4=1+4\alpha \bar{X},\; G_5=-4\alpha \ln\bar{|X|},
\end{equation}
while the metric functions and the expression for the kinetic term $\bar{X}$ are given by
\begin{equation}
    \label{eq:LP-hX-tilde}
   \bar{h}(r)=\bar{f}(r)=1+\frac{r^2}{2 \alpha}\left(1-\sqrt{1+\frac{8\alpha M}{r^3}}\right) \quad {\rm and} \quad 
   \bar{X}(r)=\frac{ 2\sqrt{\bar{f}+q^2 r^2}-1-\bar{f}}{2 r^2}\,,
\end{equation}
with $\alpha$ being a positive coupling constant.
The expression for the scalar field can be derived directly from the expression of the kinetic term by using the definition \eqref{eq:X-inhomo}. 
By doing so, one finds that
\begin{equation}
\bar{\phi}=q t+\int \diff r \; \frac{\sqrt{\bar{f}+q^2 r^2}-\bar{f}}{r \bar{f}}\,.
\end{equation}
As we showed in Sec.\,\ref{Sec:Min}, the qEGB solution possesses a nontrivial vacuum.
Even for $M=0$, for which the metric reduces to the flat Minkowski spacetime, both the kinetic term and the scalar field remain nontrivial.
The expression of the kinetic term $\bar{X}_0(r)\equiv \bar{X}(r)|_{M=0}$ in this case, is given by \eqref{eq:LP-X0}, which is a regular and strictly decreasing function for positive $r$. 
In particular, $\bar{X}_0$ acquires the finite value $q^2/2$ as $r$ approaches zero, while it vanishes at the boundary of spacetime, $r\to +\infty$.

Starting from the qEGB Minkowski vacuum with $M=0$, and performing a disformal transformation, it is straightforward to verify that the disformed metric is non-homogeneous and is generically regular for all smooth functions $D(\bar{X})$ since $g_{rr}=1$ at $r=0$.
Indeed one can always perform a rescaling of time and fix $g_{tt}$ to unity (at $r=0$ or infinity) but one can never do so for $g_{rr}$. 
The disformed metric is nevertheless not trivial anymore for $q\neq 0$ unlike the seed Minkowski metric. 
In other words, $q$ now explicitely modifies the metric not only the scalar. Generically we avoid any negative powers of $X$ in $D$ as they would spoil the asymptotics and render spacetime singular.

Let us start by supposing that $D$ is constant as we then remain in Horndeski. In this case (and whenever $D$ includes a constant term) it is easy to see that one has a solid angle deficit changing the asymptotics from Minkowski to locally Minkowski. Indeed the area of the sphere is reduced or augmented according to $D$ as, $4\pi r^2 (1+D q^2)$. As a result one can have double images of objects residing behind the black hole. This is the spacetime structure akin to the global monopole solution of Barriola and Vilenkin \cite{Barriola:1989hx}. These asymptotics persist independently of the inclusion of mass in the seed metric or not. This characterizes therefore the Horndeski disformal transformations.

We take now a linear disformal transformation $D=\xi \bar X$, with $\xi$ being a coupling constant with dimensions $({\rm length})^4$. We omit any constant as we know it would change the asymptotics.
Then, the disformed metric functions are given by
\begin{equation}
    \label{eq:dis-min}
    h(r)=1+\frac{\xi  q^4}{1+\sqrt{1+q^2 r^2}}\,,\qquad f(r)=h(r)\left[1+\frac{2\xi  q^4}{\left(1+\sqrt{1+q^2 r^2}\right)^2}\right]^{-1}\,.
\end{equation}
Two distinct cases arise here depending on the sign of the coupling constant. 
If $\xi>0$ the scalar hair $q$ gives a gravitational soliton. 
Indeed $h(r)$ is strictly positive and so is $1+2 D \bar X$.
Also, since $g_{rr}=1+\mathcal{O}(r^2)$ the metric describes a regular solution. 
However, when $\xi<0$ and scalar charge is large enough so that $1+\frac{\xi q^4}{2}<0$, then both $h(r)$ and $1+2 D \bar X$ have a root at finite $r$. 
Specifically, the root of $h$ corresponds to a horizon and occurs at greater $r$ than the root of $1+2 D \bar X$. 
At the latter root, $f(r)$ becomes discontinuous and thus spacetime ceases to be defined, and we never reach $r=0$. 
If the charge is small enough so that $1+\frac{\xi q^4}{2}>0$, we again have a solitonic solution. 
Consequently, the case $\xi<0$ is more subtle than $\xi>0$. 
The condition $1+\frac{\xi q^4}{2}$ simply corresponds to $1+2 D \bar X>0$ for all $r$, which effectively guarantees that the disformal transformation is invertible in the permitted $r$-region and hence smooth.

\begin{figure}[t]
    \centering
    \includegraphics[width=0.5\textwidth]{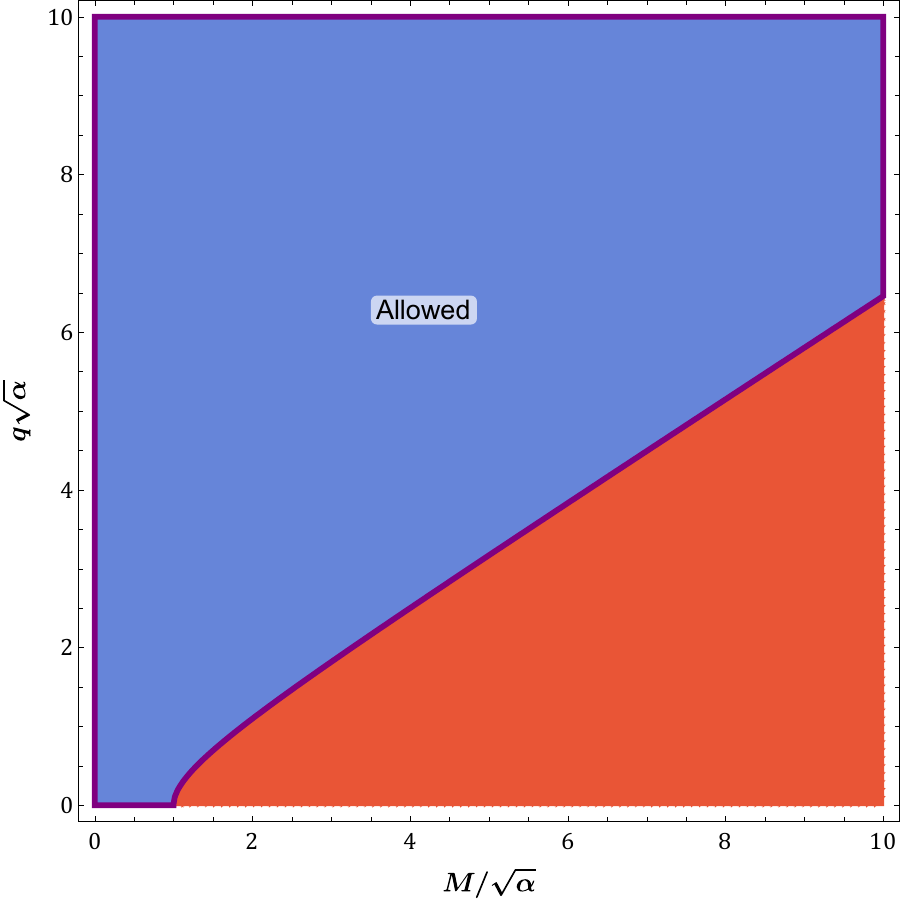}
    \vspace*{-0.5em}
    \caption{Parameter space spanned by the dimensionless parameters $M/\sqrt{\alpha}$ and $q\sqrt{\alpha}$. The upper blue region indicates the permissible range of parameter values.}
    \label{fig:par-L}
\end{figure}

Before we examine the solutions arising from the disformal transformation $D=\xi \bar{X}$ with $M\neq 0$, it is essential to first determine the region of parameter space for which the kinetic term $\bar{X}$ remains real and well-defined for all positive $r$.
This requirement is crucial because the model functions of the disformed metric explicitly depend on the seed kinetic term $\bar{X}$. While $\bar{X}$ is guaranteed to be real within the causal region of the seed spacetime,\,\footnote{The horizon radius $r_h$ of the seed qEGB metric is the root of the equation $\bar{h}(r_h)=\bar{f}(r_h)=0$, which from \eqref{eq:LP-hX-tilde} can be determined to be $r_h=M+\sqrt{M^2-\alpha}$. Hence, as long as $M^2>\alpha$, the quantity under the square root in the expression of $\bar{X}$ will be positive-definite for all $r>r_h$.} this property is not automatically preserved under the disformal transformation.
One can readily find values for the parameters $M$, $q$, and $\alpha$ for which $\bar{X}$ becomes complex-valued at finite $r$, causing the disformed metric functions to turn complex well before reaching $r=0$. 
Since such a behavior is unphysical, it is necessary to determine the region of parameter space ($M,q,\alpha)$ for which both $\bar{X}$ and the target metric of the beyond Horndeski theory remains real-valued for all $r$. 
To this end, it suffices to examine the quantity $\bar{f}+q^2r^2$, which appears under the square root in $\bar{X}$.
The sign of $\bar{f}+q^2 r^2$ determines whether the kinetic term of the seed solution takes complex values or not.
It is straightforward to see that $\bar{f}+q^2 r^2$ can be written in the form
\begin{equation}
    \label{eq: xcon}
    \bar{f}+q^2 r^2=1+\frac{\breve{r}^2}{2}\left(1+2\breve{q}^2-\sqrt{1+\frac{8\breve{M}}{\breve{r}^3}} \right)\,,
\end{equation}
where $\breve{r}\equiv r/\sqrt{\alpha}$, $\breve{q}\equiv q\,\sqrt{\alpha}$, and $\breve{M}\equiv M/\sqrt{\alpha}$. 
Notice that all quantities have been rescaled relative to the coupling constant $\alpha$, ensuring that our conclusions will remain valid independently of $\alpha$.
Turning to Eq.\,\eqref{eq: xcon}, it is clear that for $\breve{M}>0$ the expression inside the brackets will inevitably become negative as the radial coordinate approaches zero.
Consequently, to keep $\bar{f}+q^2 r^2$ positive, $\breve{M}$ and $\breve{q}$ must be chosen such that the bracketed term becomes negative for sufficiently small values of $\breve{r}$, so that the overall contribution is suppressed by the $\breve{r}^2$ factor.
To identify the allowed values of the parameters $\breve{M}$ and $\breve{q}$ that ensure the positivity of $\bar{f}+q^2 r^2$, we first need to make a few key observations. 
Firstly, at $\breve{r}\rightarrow 0$, the r.h.s. of equation \eqref{eq: xcon} becomes unity, while at the boundaries of spacetime, where $\breve{r}\rightarrow +\infty$, the function diverges to infinity.
This implies that if the function becomes negative definite at any point, it will only do so temporarily since its values at both ends of its domain are positive.
Therefore, if the function takes negative values for certain values of the parameters $\breve{M}$ and $\breve{q}$, it is guaranteed to have a global minimum. 
By identifying the location of this minimum and requiring the value of the function at this point to be positive, we can ensure that the function remains positive over its entire domain. 
This approach allows us to determine the permissible values of the parameters $\breve{M}$ and $\breve{q}$.
In Fig. \ref{fig:par-L} we depict the allowed values of the dimensionless quantities $\breve{M}=M/\sqrt{\alpha}$ and $\breve{q}=q \sqrt{a}$ for which the function $\bar{f}+q^2r^2$ remains positive over its entire domain and thus the kinetic term $\bar{X}$ remains real-valued.
It is interesting to notice that in the case, where $M/\sqrt{\alpha}<1$, the kinetic term $\bar{X}$ will be positive definite regardless of the value of the scalar charge $q$.
This is not a coincidence, since for $M^2<\alpha$ the seed qEGB metric describes a naked singularity with the metric functions being regular with an expansion at $r\to 0$ of the form
$$\bar{h}(r)=\bar{f}(r)\xrightarrow{r\rightarrow 0^+} 1-2\sqrt{\frac{M r}{\alpha^2}}+\mathcal{O}(r^2)\,,$$
and no horizons are present.

\begin{figure}[t]
    \centering
    \begin{subfigure}[b]{0.47\textwidth}
    \includegraphics[width=1\textwidth]{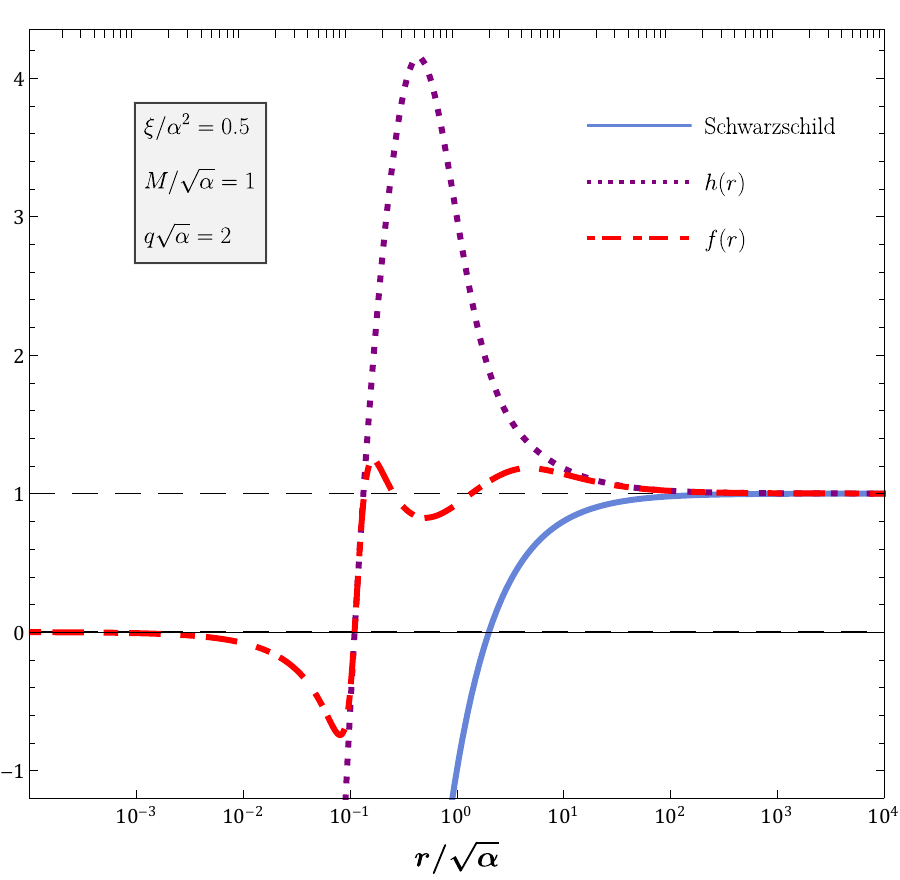}
    \caption{\hspace*{-0.5em}}
    \label{subf:lin-dis1}
    \end{subfigure}
    \hfill
    \begin{subfigure}[b]{0.478\textwidth}
    \includegraphics[width=1\textwidth]{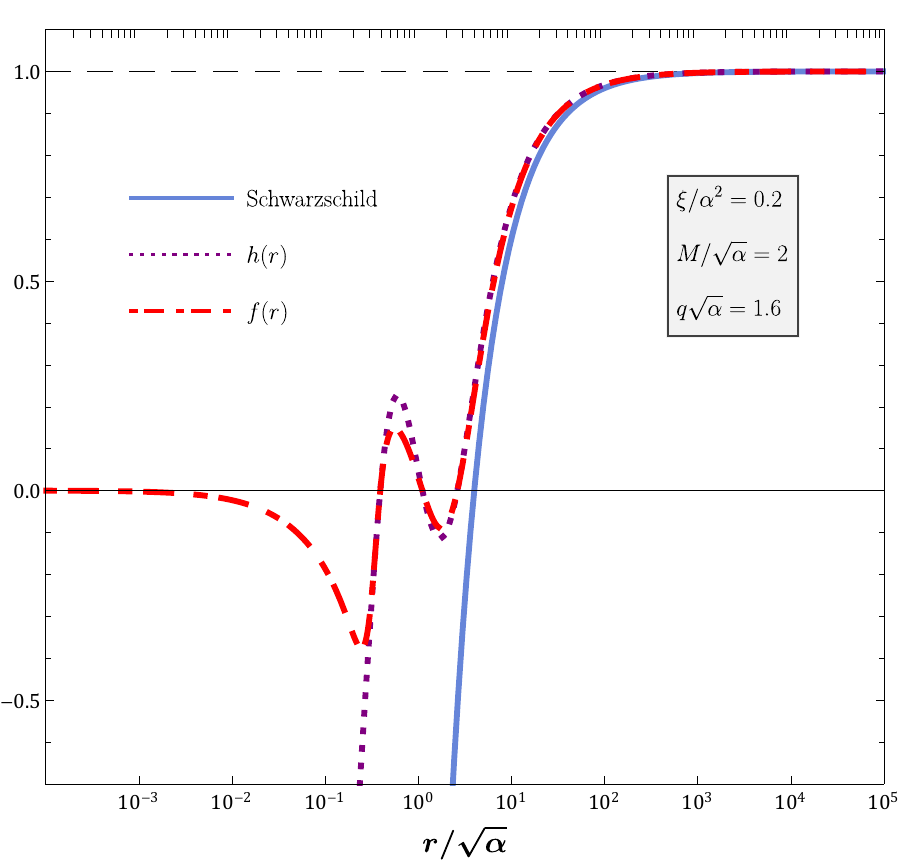}
    \caption{\hspace*{-1em}}
    \label{subf:lin-dis2}
    \end{subfigure}
    \vspace*{-0.5em}
    \caption{The disformed metric functions $h(r)$ and $f(r)$ for two different sets of the dimensionaless parameters $\{\xi/\alpha^2,M/\sqrt{\alpha},q\sqrt{\alpha} \}$. (a) Single-horizon black hole, (b) black hole with multiple horizons.}
    \label{fig:lin-dis}
\end{figure}


Let us now consider the disformal transformation described by the linear disformal factor $D(\bar{X})=\xi \bar{X}$, for the seed metric \eqref{eq:LP-hX-tilde} and $M\neq 0$.
For all values in the blue region of Fig.\,\ref{fig:par-L} and for $\xi>0$, the disformal transformation is invertible since
\begin{equation}
    1+2\bar{X}D(\bar{X})=1+2\xi \bar{X}^2>0\,.
\end{equation}
In Fig. \ref{fig:lin-dis}, we depict the profiles of the metric functions $h(r)$ and $f(r)$, for two different sets of the dimensionless parameters $\{\xi/\alpha^2,M/\sqrt{\alpha},q\sqrt{\alpha} \}$, for $\{0.5,1,2\}$ in Fig.\,\ref{subf:lin-dis1}, and for $\{0.2,2,1.6\}$ in Fig.\,\ref{subf:lin-dis2}.
We observe, that the disformed solution may describe black holes with one or multiple horizons. 
In both scenarios though, the resulting black holes are more compact than the Schwarzschild solution since the event horizon radii of the former solutions are smaller compared to that of the Schwarzschild black hole of the same mass $M$.
The explicit expressions for the metric functions can be determined from Eqs. \eqref{eq:dis-metr} and \eqref{eq:LP-hX-tilde}.
Below, we present the expansion of the metric functions at the boundaries of spacetime:
\begin{align}
    \label{hL}
    &h(r)\xrightarrow{r\rightarrow +\infty}1-\frac{2M-q^3\xi}{r}-\frac{q^2\xi}{r^2}+\mathcal{O}(1/r^3)\,,\\[2mm]
    \label{fL}
    &f(r)\xrightarrow{r\rightarrow +\infty}1-\frac{2M-q^3\xi}{r}-\frac{3q^2\xi}{r^2}+\mathcal{O}(1/r^3)\,.
\end{align}
We notice that asymptotically the black hole resembles the Reissner-Nordstr\"{o}m (RN) spacetime, with the scalar hair $q$ playing the role of the electric charge.
However, in this case, the spacetime is inhomogeneous and the signs in $\mathcal{O}(1/r^2)$ terms are flipped, indicating an attractive contribution of the scalar hair.
Additionally, we see that the scalar hair appears also in order $\mathcal{O}(1/r)$ a characteristic that is absent for RN black holes.
Consequently, the ADM mass of the disformed solution depends on both the ADM mass of the seed solution $M$ and the scalar hair $q$ that is acquired after the transformation.
Setting $M=0$, the solution reduces to that of Eqs.\,\eqref{eq:dis-min}, describing solitonic solutions.


\section{Concluding remarks}

In this work, we have addressed a basic but consequential question for scalar-tensor gravity: which shift-symmetric Horndeski theories legitimately admit the simple Minkowski and de Sitter vacua? 
By imposing consistency conditions directly on the field equations rather than on observational grounds, we derived clear selection rules that single out the subclasses of Horndeski models capable of supporting nontrivial Minkowski and de Sitter backgrounds. 
In doing so, we introduced and characterized the notion of stealth vacua, namely configurations in which the scalar field is active but leaves the vacuum metric unaltered, and showed how this concept refines the standard notion of vacuum in scalar-tensor extensions of GR.

The filtering mechanism that emerges from our analysis has two practical payoffs. 
First, it excludes broad classes of the theory space that produce pathological or physically uninteresting vacua. 
Second, it isolates a tractable set of models in which the usual toolkit of perturbation theory, gravitational-wave calculations, and cosmological perturbations may be applied without ambiguity. 
In short, requiring the existence of well-defined vacua provides a  criterion for theory selection that complements, and in some cases precedes, phenomenological constraints.

Building on this, the second half of the paper focuses on constructing compact-object solutions in beyond-Horndeski gravity.
Starting from the regularized qEGB black hole \cite{Charmousis:2019vnf} as a seed solution in Horndeski gravity, we performed a linear disformal transformation to generate exact, nonhomogeneous solutions in the corresponding/target beyond-Horndeski theory. 
The map produces both solitonic spacetimes sourced by nontrivial scalar profiles on a seed Minkowski spacetime, and black holes that carry primary scalar hair in the target theory.
These explicit constructions demonstrate that disformal relations can both preserve and qualitatively change the physical character of solutions, promoting stealth-like configurations to compact objects with independent scalar charges.

We also identified the conditions under which the disformal map is well defined and the transformed geometries remain physically acceptable. 
These technical constraints, invertibility of the transformation, avoidance of curvature singularities introduced by the map, and the maintenance of appropriate asymptotics, delineate the region where the new solutions are reliable. 
The natural next steps are a careful analysis of linear and nonlinear stability, extensions to rotating and dynamical spacetimes, the inclusion of matter couplings, and targeted phenomenological studies (gravitational-waveforms, quasi-normal mode spectra, black-hole imaging and accretion signatures, and cosmological implications) that can bridge the gap between these theoretical constructions and observations.

Overall, the paper offers a twofold contribution. 
On the one hand, it supplies a simple, physically motivated criterion to pare down the vast landscape of shift-symmetric Horndeski theories to those that admit well-behaved vacua. 
On the other hand, it provides explicit methods to generate nontrivial compact-object solutions in beyond-Horndeski gravity, thereby expanding the catalogue of spacetimes that must be confronted by experiment and observation. 
Together, these results constitute a practical framework for both theory building and phenomenology. 
They show how demanding the existence of ``canonical vacua" sharply constrains model space, and how disformal mappings can be exploited to explore the rich solution structure that lies beyond Horndeski gravity.

It would be interesting to investigate how the nontrivial vacua we have encountered here can be extended or modified for other modified gravity theories such as scalar-vector-tensor theories. Such theories acquiring a nontrivial vectorial sector can acquire interesting dark energy properties \cite{Mironov:2024idn, Mironov:2024wbx, Babichev:2024kfo, Mironov:2025dzz} if they maintain $U(1)$ symmetry. More recently they have been shown to have primary hair black holes \cite{Charmousis:2025xug} when they are of Proca type. These questions amongst others may be interesting for future study.

\section*{Acknowledgements}
We would like to thank Mokhtar Hassaine and Karim Noui for interesting discussions. C.C. acknowledges partial support of ANR grant StronG (ANR-22-CE31-0015-01). T.N. is supported by
IBS under the project code IBS-R018-D3.


\bibliography{Refs.bib}

\end{document}